\documentclass[a4paper,11pt]{article}
\usepackage{pos}

\title{Spin Physics Detector project at JINR}
\author*[a]{A. Guskov on behalf of the SPD collaboration}

\affiliation[a]{Joint Institute for Nuclear Research,\\
  Joliot-Curie 6, Dubna, Russia}

\emailAdd{avg@jinr.ru}

\abstract{The Spin Physics Detector at the constructing NICA collider (JINR, Dubna) is a universal facility to investigate the spin structure of the proton and deuteron and the other spin-related phenomena with polarized proton and deuteron beams at a collision energy up to 27 GeV. A comprehensive study of the unpolarized and polarized gluon content of the nucleon at large Bjorken-x using different complementary probes such as charmonia, open charm, and prompt photon production processes is the central point of the SPD physics program. In the polarized proton-proton collisions, the SPD experiment at NICA will cover the kinematic gap between the low-energy measurements at ANKE-COSY and SATURNE and the high-energy measurements at the Relativistic Heavy Ion Collider, as well as the planned fixed-target experiments at the LHC. The possibility for NICA to operate with polarized deuteron beams at such energies is unique. The SPD experimental setup is planned as a multipurpose universal 4$\pi$ detector with advanced tracking and particle identification capabilities, electromagnetic calorimeter, and muon (range) system. To minimize possible systematic effects, SPD will be equipped with a free-running data acquisition system. The spin physics program at the SPD is expected to start after 2025 and to extend for about 10 years.}

\FullConference{%
  *** Particles and Nuclei International Conference - PANIC2021 ***\\
  *** 5 - 10 September, 2021 ***\\
  *** Online ***
}

\begin{document}
\maketitle

\section{Introduction}
Quantum chromodynamics has remarkable success in describing the high-energy and large-momentum transfer processes, where quarks and gluons that are the fundamental constituents of hadrons, behave, to some extent, as free particles and, therefore, the perturbative QCD approach can be used. The cross-section of a process in QCD is factorized  into two parts: the process-dependent perturbatively-calculable short-distance partonic cross-section (the hard part) and universal long-distance functions, PDFs, and FFs (the soft part). The parton distributions could be applied also to describe the spin structure of the nucleon that  is built up from the intrinsic spin of the valence and sea quarks (spin-1/2), gluons (spin-1), and their orbital angular momenta. A full description can be given in terms of the so-called transverse-momentum dependent parton distribution functions (TMD PDFs).
Notwithstanding, the progress achieved during the last decades in the understanding of the quark contribution to the nucleon spin, the gluon sector is much less developed. One of the difficulties is the lack of direct probes to access gluon content in high-energy processes. 
While the quark contribution to the nucleon spin was determined quite precisely in semi-inclusive deep-inelastic scattering (SIDIS) experiments like EMC, HERMES, and COMPASS, the gluon contribution is still not well-constrained even so it is expected to be significant. 

Nevertheless, the largest fraction of hadronic interactions involves  low-momentum transfer processes in which the effective strong coupling constant is large and the description within a perturbative approach is not adequate. A number of (semi-)phenomenological approaches have been developed through the years to describe strong interaction in the non-perturbative domain starting from the very basic principles. They successfully describe such crucial phenomena, as the nuclear properties and interactions, hadronic spectra,  deconfinement, various polarized and unpolarized effects in hadronic interaction, etc. The transition between the perturbative and non-perturbative QCD is also a subject of special attention.  In spite of a large set of experimental data and huge experience in few-GeV region with fixed-target experiments worldwide, this energy range still attracts both experimentalists and theoreticians. 
\begin{figure}[h!]
  \begin{center}
    \includegraphics[width=0.75\textwidth]{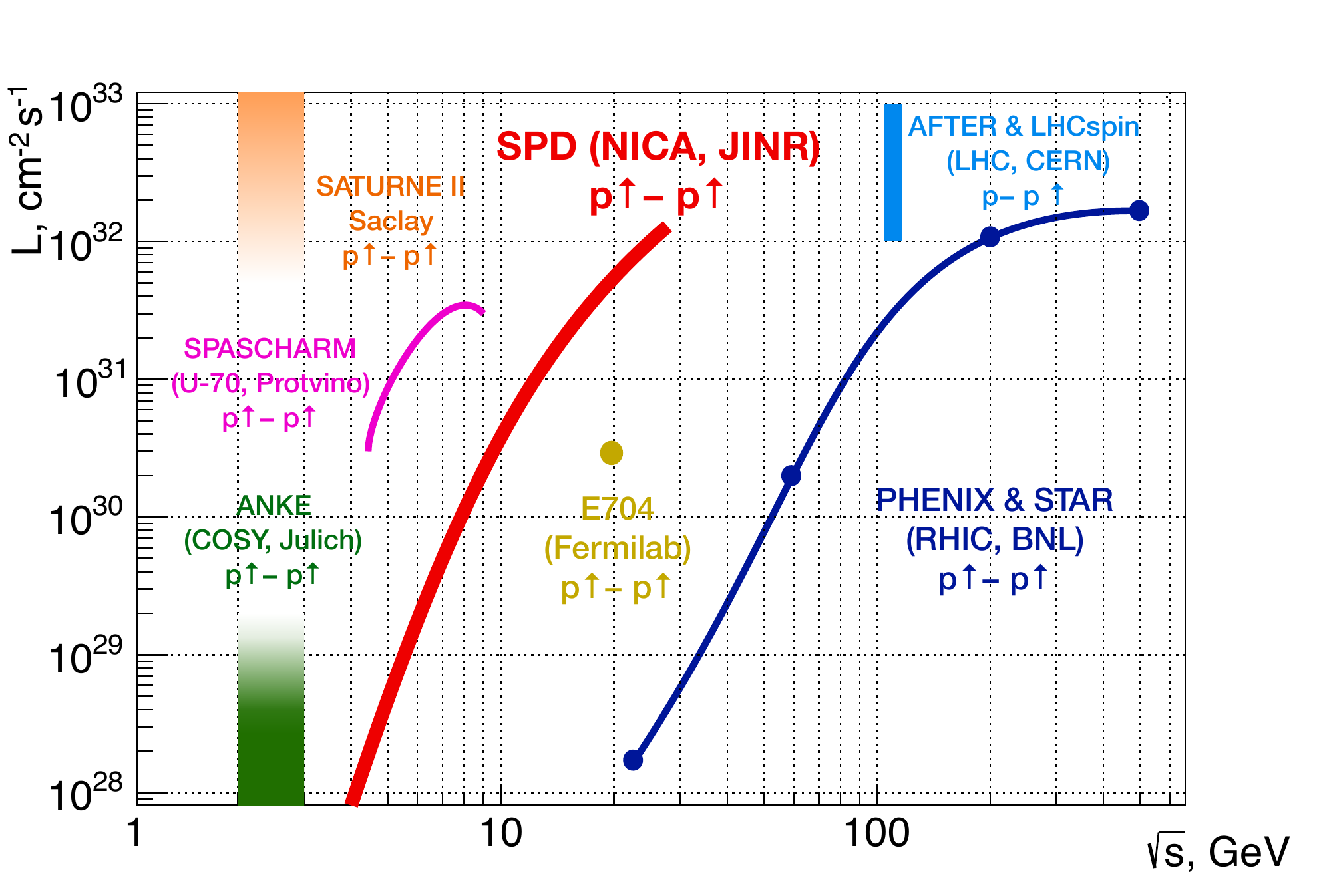}
  \end{center}
  \caption{NICA SPD and the other past, present, and future experiments with polarized protons.}
  \label{S_L}
\end{figure}

The Spin Physics Detector is a planned experimental setup at the NICA collider that is under construction at the Joint Institute for Nuclear Research (Dubna, Russia)  is intended to study the spin structure of the proton and deuteron and the other spin-related phenomena with polarized proton and deuteron beams at a collision energy up to 27 GeV and a luminosity up to $10^{32}$ cm$^{-2}$ s$^{-1}$. In the polarized proton-proton collisions, the SPD experiment \cite{SPDproto:2021hnm, spd_url} at NICA will cover the kinematic gap between the low-energy measurements at ANKE-COSY and SATURNE and the high-energy measurements at the Relativistic Heavy Ion Collider, as well as the planned fixed-target experiments at the LHC (see Fig.  \ref{S_L}). The possibility for NICA to operate with polarized deuteron beams at such energies is unique. 

\section{SPD physics goals}
 SPD is planned to operate as a universal facility for  comprehensive tests of the basics of the QCD. The main goal of the experiment is the study of the unpolarized and polarized gluon content of the proton at large Bjorken-$x$, using different complementary probes such as charmonia, open charm, and prompt photon production processes \cite{Arbuzov:2020cqg}. 
 The experiment aims at providing access to the gluon helicity, gluon Sivers, Boer-Mulders functions, and other TMD PDFs in the proton via the measurement of specific single and double spin asymmetries. The kinematic region to be covered by SPD (Fig. \ref{SPD-LA_EX}) is unique and has never been accessed purposefully in polarized hadronic collisions. Quark TMD PDFs, as well as spin-dependent fragmentation functions could also be studied.  The results expected to be obtained by SPD will play an important role in the general understanding of the nucleon gluon content and will serve as a complementary input to the ongoing and planned studies at RHIC, and future measurements at the EIC (BNL) and fixed-target facilities at the LHC (CERN). Simultaneous measurement of the same quantities using different processes at the same experimental setup is of key importance for the minimization of possible systematic effects. 
 
 The naive model describes the deuteron as a weakly-bound state of a proton and a neutron mainly in S-state with a small admixture of the D-state. However, such a simplified picture failed to describe the HERMES experimental results on the $b_1$ structure function. A unique possibility to operate with polarized deuteron beams brings us to the world of the tensor structure of the deuteron.  A possible non-baryonic content in the deuteron could be accessed via the measurement of the gluon transversity distribution and the comparison of the unpolarized gluon PDFs in the nucleon and deuteron at high values of $x$. 

 SPD has an extensive physics program for the first stage of the NICA collider operation with reduced luminosity and collision energy of the proton and ion beams, devoted to comprehensive tests of the various phenomenological models in the non-perturbative and transitional kinematic domain. It includes such topics as the spin effects in elastic scattering, exclusive reactions as well as in hyperons production, multiquark correlations and dibaryon resonances, charmonia and open charm production, physics of light and intermediate nuclei collision, hypernuclei, etc. \cite{Abramov:2021aey}.
 
 The proposed physics program covers up to 5 years of the NICA collider running.

 \begin{figure}[!h]
  \begin{center}
    \includegraphics[width=0.75\textwidth]{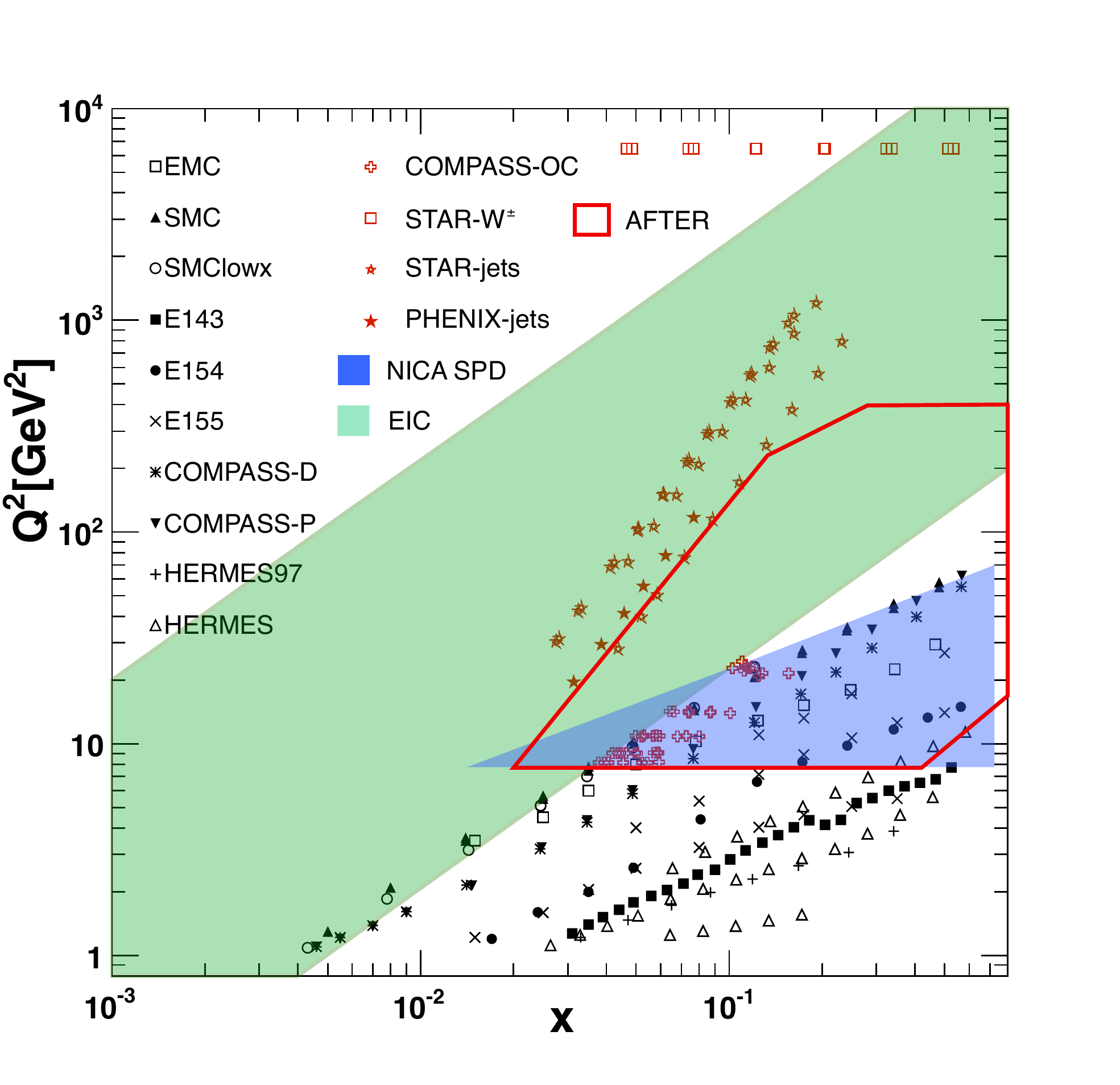}
  \end{center}
  \caption{Kinematic coverage of SPD in the charmonia, open charm, and prompt photon production processes.}
  \label{SPD-LA_EX}
\end{figure}

\section{SPD experimental setup}

 \begin{figure}[!h]
  \begin{center}
    \includegraphics[width=0.75\textwidth]{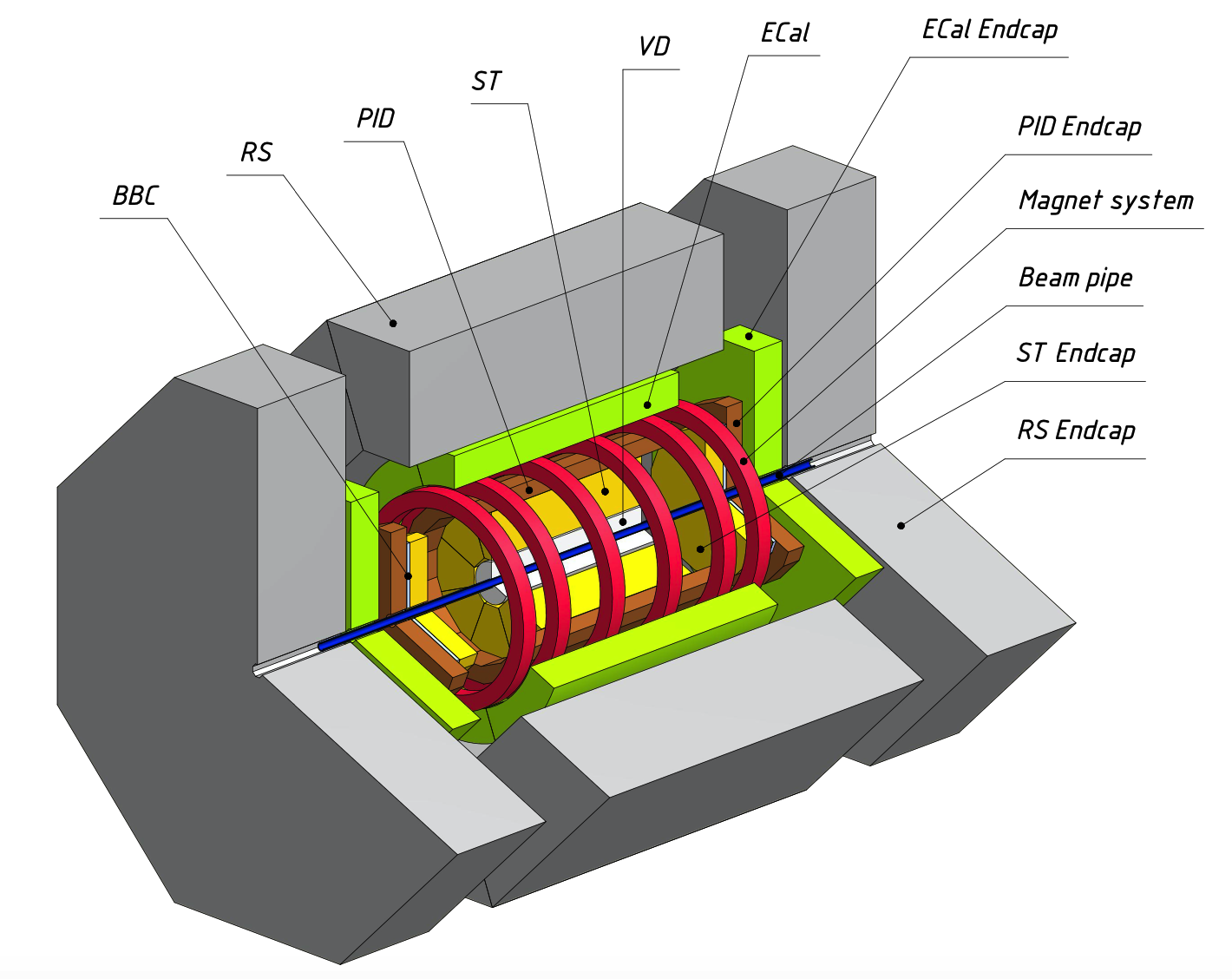}
  \end{center}
  \caption{General layout of the SPD setup.}
  \label{SPD-LA_EX}
\end{figure}
The physics goals dictate the layout of the detector. The SPD experimental setup is being designed as a universal $4\pi$ detector with advanced tracking and particle identification capabilities based on modern technologies. The silicon vertex detector (VD) will provide resolution for the vertex position on the level of below 100 $\mu$m needed for reconstruction of secondary vertices of $D$-meson decays. The straw tube-based tracking system (ST) placed within a solenoidal magnetic field of up to 1 T at the detector axis should provide the transverse momentum resolution $\sigma_{p_T}/p_T\approx 2\%$ for a particle momentum of 1 GeV/$c$. The time-of-flight system (PID) with a time resolution of about 60 ps will provide $3\sigma$ $\pi/K$ and $K/p$ separation of up to about 1.2 GeV/$c$ and 2.2 GeV/$c$, respectively. Possible use of the aerogel-based Cherenkov detector could extend this range. Detection of photons will be provided by the sampling electromagnetic calorimeter (ECal) with the energy resolution $\sim 5\%/\sqrt{E}$. To minimize multiple scattering and photon conversion effects for photons, the detector material will be kept to a minimum throughout the internal part of the detector. The muon (range) system (RS) is planned for muon identification. It can also act as a rough hadron calorimeter. The pair of beam-beam counters (BBC) and zero-degree calorimeters will be responsible for the local polarimetry and luminosity control.  To minimize possible systematic effects, SPD will be equipped with a triggerless DAQ system. A high collision rate (up to 4 MHz) and a few hundred thousand detector channels pose a significant challenge to the DAQ,  online monitoring, offline computing system, and data processing software.

\section{Summary}
The proposed physics program and the concept of the Spin Physics Detector facility have been proposed to be performed at the NICA collider at JINR. The measurements at SPD have bright perspectives to make a unique contribution and challenge our understanding of the spin structure of the nucleon and the nature of the strong interaction. We are open to exciting and challenging ideas from theorists and experimentalists worldwide.

\end{document}